# The nonequilibrium back-reaction of Hawking radiation to a Schwarzschild black hole

He Wang[1,2], Jin Wang[3*]

[1]College of Physics, Jilin University, Changchun, 130021

[2] State Key Laboratory of Electroanalytical Chemistry, Changchun Institute of Applied Chemistry, Changchun, 130021

[3]Department of Chemistry and Physics, Stony Brook University, Stony Brook, NY 11794-3400

[*]**Corresponding Author:** jin.d.wang@gmail.com

## Abstract

We investigate the nonequilibrium back reaction on the Schwarzschild black hole from the radiation field. The back reactions are characterized by the membrane close to the black hole. When the membrane is thin, we found that larger temperature difference can lead to more significant negative surface tension, larger thermodynamic dissipation cost and back reaction in energy and entropy as well as larger black hole area. This may be relevant to the primordial black holes in early universe. Moreover, our nonequilibrium model can resolve the inconsistency issue of the black hole back reaction under zero mass limit in the equilibrium case. In the thick membrane case, the nonequilibrium back reaction is found to be more significant than that in the thin membrane case. The nonequilibrium temperature difference can increase the energy and entropy loss as well as the thermodynamic dissipation of the black hole and the membrane back reactions. The nonequilibrium dissipation cost characterized by the entropy production rate appears to be significant compared to the entropy rate radiated by the black hole under finite temperature difference. This may shed light on the black hole information paradox due to the information loss from the entropy production rate in the nonequilibirum cases. The nonequilibrium thermodynamic fluctuations can also reflect the effects of the back-reactions of the Hawking radiation on the evolution of a black hole.

# I. Introduction

Hawking radiation has been suggested to emerge near the black hole event horizon with a thermal spectrum.[1] The radiated particles carry energy-momentum so that the radiation can have a back-reaction on the black hole. One way of quantifying this back reaction is to find a rational renormalized energy-momentum tensor$\langle 0|T_{\mu\nu}|0\rangle$ and regard it as a source. In this way, one can study the back reaction by solving the semi-classical Einstein equation approximatively as $G_{\mu\nu} = -k\langle 0|T_{\mu\nu}|0\rangle$.[2] The back-reaction and/or quantum corrections will modify Hawking temperature and entropy. (quantum corrections and back reaction are often intertwined)[3][4][5][6]. However, we often have poor knowledge about the source $\langle 0|T_{\mu\nu}|0\rangle$. Therefore, this dynamical approach is often limited. On the other hand, thermodynamic approach can be used to study the black hole back-reaction problem without the direct use of $\langle 0|T_{\mu\nu}|0\rangle$. The thermodynamic approach has been applied to quantify the back reaction to Schwarzschild black hole [7][8], Kerr black hole and Kerr-Newmann black hole.[9][10]

Both dynamical and thermodynamic approaches are under a precondition: black hole is in equilibrium with the radiation. However, due to a negative thermal capacity, it is difficult for black hole to maintain the equilibrium unless the black hole is in the box. A negative thermal capacity can magnify the small temperature fluctuations, leading to thermodynamic instability.

In this study, we investigate the nonequilibrium back reaction of the radiation field on the Schwarzschild black hole by considering both the thin membrane and thick membrane cases. The nonequilibriumness is characterized by the temperature difference between the black hole and the radiation field, ΔT. We found that in general the nonequilibrium effects characterized by the temperature difference can increase the energy and entropy loss as well as the entropy production rate of the black hole and the membrane back reactions. We uncover that the nonequilibrium back-reaction can lead to significant changes in the energy, entropy of some primordial black holes created in the early universe for the thin membrane. We also found that larger temperature difference can lead to more significant negative surface tension and back reaction as well as larger black hole area. Moreover, our nonequilibrium model can resolve the inconsistency issue of the black hole back reaction under zero mass limit in the equilibrium case. As the mass of the black hole $M_{\text{BH}} \to 0$, the equilibrium dressed energy and entropy of the black hole do not appear to vanish. By considering the nonequilibrium scenario in the present case, we found that as $M_{\text{BH}} \to 0$, both $M_{\text{dressed}}$ and $S_{\text{dressed}}$ will vanish. This provides another way of resolving the issue of inconsistency of the black hole back reaction under zero mass limit without using the quantum correction. This leads to the conclusion that our nonequilibrium model is more natural and physically reasonable than the equilibrium model for the black hole back reaction.

On the other hand, the associated Hawking temperature $T = \frac{1}{8\pi M_{BH}}$ is divergent at zero black hole mass or size. Both the equilibrium and the nonequilibrium scenarios have a temperature divergence with or without the back reaction. Due to the quantum uncertainty principle, the size of the black hole cannot exactly approach to zero. This is bounded by the large energy/energy momentum fluctuations at the small black hole size. Therefore, one can avoid such singularity point where the radius of the black hole approaches zero (r → 0) and the associated temperature approaches infinity ($T$ → 0)when considering quantum effects. The quantum corrections can give rise to well behaved temperature and entropy in the $M_{BH}$ → 0 limit[3].For the thick membrane case, we consider the case where the temperature varies uniformly in the membrane. The back reaction from the thick membrane is found to be more significant compared to that in the thin membrane case. Importantly, we quantified the nonequilibrium thermodynamic dissipation represented by the entropy production rate for both the thin membrane and thick membrane which increases significantly as the temperature difference increases. We show that once there is a finite temperature difference, the nonequilibrium entropy production rate appears to be significant compared to the entropy rate radiated by the black hole. The nonequilibrium entropy production rate can thus have an impact on the black hole evaporation rate. This may be significant for black hole information paradox due to the information loss by the entropy production rate originated from the nonequilibriumness characterized by the temperature difference between the black hole and the radiation environment. We explored the nonequilibrium thermodynamic fluctuations of the black hole, which can reflect the effects of the back-reactions of the Hawking radiation on the evolution of a black hole evaporation.

In section **II**, we firstly developed a thermodynamic model for studying back reaction, and then we mainly focus on the thin membrane case. In section **III**, we consider the thick membrane case when the temperature varies uniformly within the membrane. We found the back reaction to the black hole mass, entropy and the nonequilibrium thermodynamic cost characterized by the entropy production rate can be significant compared to that of the black hole and its evaporation rate respectively under the finite temperature difference. In section **IV**, nonequilibrium thermodynamic fluctuations of the black hole is studied and we found the more massive the black hole is, the weaker its back-reaction is. In section **V**, we give a conclusion for our study.

## II. Back reactions to the black holes: thin membrane case

### 1. Treat the back reaction as a membrane: equilibrium scenario

In this subsection, we briefly review the thermodynamic approach to back reaction on Schwarzschild black hole at equilibrium state [7][8]. Let us consider a

Schwarzschild black hole in a large adiabatic bulk volume filled with thermal radiation field, as shown in Figure 1. One can approximately assume that the radiation field is filled in the whole universe and its temperature is unchanged during the dynamical processes. The bulk is connected to a piston in the asymptotically flat regime by a narrow tube. The piston can do work on the radiation field.

The total volume change of the radiation field can be described as $\delta V = \delta V_{BH} + \delta V_{ex}$. $\delta V_{BH}$ represents the volume change of the black hole and $\delta V_{ex}$ represents the volume change of the radiation field. Then, the total energy and entropy of the system including the back reaction are $E = M_{BH} + M_R + M_H$ and $S = S_{BH} + S_R + S_H$ respectively. $M_{BH}$ represents the mass of black hole, which is also the energy of black hole in the natural units. $M_R = aT^4 V$ represents the energy of radiation field where a represents a positive definite coefficient and $T = \frac{1}{8\pi M_{BH}}$ represents the temperature of both the black hole and the radiation field. $S_{BH} = 4\pi M_{BH}^2$ represents the entropy of black hole and $S_R = \frac{4}{3}aT^3 V$ represents the entropy of the radiation field where $V$ is the volume of the radiation field. $M_H$ and $S_H$ represent the energy and entropy of back reaction respectively.

From the above definitions and analysis, one can easily obtain that the change of the energy to be:

$$dE = dM_H - TdS_H - \frac{1}{3}aT^4 dV = dM_H - TdS_H - \frac{1}{3}aT^4 dV_{BH} - \frac{1}{3}aT^4 dV_{ex}$$

The above expression is obtained since the system is assumed to be in an adiabatic bulk. The first law of thermodynamics for such a system reads $dE = -\frac{1}{3}aT^4 dV_{ex}$, so

$$dM_H = TdS_H + \frac{1}{3}aT^4 dV_{BH}.$$

The back reaction of the radiation field to the black hole can be described by a 2 dimensional membrane with energy $M_H$ and temperature $T$. The rational is as follows. The first law of thermodynamics of the membrane reads

$$dM_H = TdS_H + \sigma dA$$

where A represents the area of the membrane and $\sigma$ represents the surface tension of the membrane. If $\sigma dA = \frac{1}{3}aT^4 dV_{BH}$, a natural inference is that the back reaction can be viewed as a membrane with temperature $T$. Therefore, to investigate the back reaction to Schwarzschild black hole from the thermal radiation under thermal equilibrium in the curved spacetime, one can consider a thermodynamic system composed of Schwarzschild black hole, thermal radiation and an effective two-dimensional thermodynamic membrane near the event horizon in flat spacetime. The membrane is formed by and physically equivalent to the back-reaction.

Let us assume that the membrane is near to the event horizon, therefore A also represents the area of the horizon, thus $\sigma = \frac{\varepsilon a}{24\pi} T^3$ and $\varepsilon$ is a dimensionless constant.

The two sides of the membrane are connected to the black hole and radiation field respectively. Moreover, this viewpoint will greatly simplify the thermodynamics of curved spacetime. Now the system contains three subsystems: the black hole, the membrane attached to horizon and the radiation field, so that one can easily derive the energy and entropy of the membrane (or the back reaction) by using thermodynamic relations. The free energy of the membrane is given as $F_H = M_H - TS_H$, whose differential form reads as

$$dF_H = -S_H dT + \sigma dA.$$

From the differential expressions of the energy and free energy from the above expression with respect to the membrane, one gets

$$M_H = \frac{4\varepsilon}{3}\frac{a}{(8\pi)^3}M_{BH}^{-1} + C_E, \quad S_H = 2\varepsilon \frac{a}{(8\pi)^2} + C_S$$

$C_E$ and $C_S$ are two integral constants. The energy and entropy of black hole with back reaction become

$$M_{\text{dressed}} = M_{BH}[1 + \frac{4\varepsilon}{3}\frac{a}{(8\pi)^3}M_{BH}^{-2}] + C_E$$

$$S_{\text{dressed}} = 4\pi M_{BH}^2(1 + 4\varepsilon \frac{a}{(8\pi)^3}M_{BH}^{-2}) + C_S$$

respectively. At certain condition (choose the proper integration constant, i.e. $C_E = 0$ and $\varepsilon = 3C_0$), the above expressions for the energy and entropy considering back reactions exactly correspond to the result of York's.[2] There are two masses now: a naked black hole mass (without back reaction) $M_{BH}$ and a dressed black hole mass (with back reaction) $M_{dressed} = M_{BH} + M_H$.

Now the temperature ambiguity may appear. Which one of the above masses is related to the equilibrium temperature of the system or further, the temperature of the black hole? Ref [7] shows that only the naked mass of the black hole determines the equilibrium temperature of the system. According to $T = \frac{1}{8\pi M_{BH}}$, one can rewrite it as

$$T \approx \frac{1}{8\pi M_{\text{dressed}}}[1 + \frac{4\varepsilon}{3}\frac{a}{(8\pi)^3}M_{\text{dressed}}^{-2}]$$

If $\varepsilon = \frac{(k_0+12)\pi^2}{10}a^{-1}$, it is exactly what York obtained [2]

$$T = \frac{1}{8\pi M_{\text{dressed}}}[1 + \frac{k_0+12}{3840\pi}M_{\text{dressed}}^{-2}]$$

We have reviewed the discussions on the effectiveness of thermodynamics approach to back reaction. The main idea is to consider back reaction as a membrane attached to black hole and study the thermodynamics of the system, rather than directly calculating through the semiclassical Einstein equation $G_{\mu\nu} = -k\langle 0|T_{\mu\nu}|0\rangle$. In other words, one can just study the back reaction of the black hole through an effective membrane in flat Minkowski spacetime without considering the gravity or curved space time explicitly, giving equivalent thermodynamics. However, if we take a limit $M_{BH} \to 0$, it appear that $M_{\text{dressed}}$ and $S_{\text{dressed}}$ do not vanish. This creates an obvious inconsistency issue here with the non-zero back reaction when the black hole is not present any more. In what follows, we show that this inconsistency issue of the black hole back reaction under zero mass limit in the equilibrium scenario can be resolved by considering nonequilibrium effect. Therefore, a natural question becomes then: what is the effect of the membrane (back reaction) in nonequilibrium case between the black hole and the radiation field?

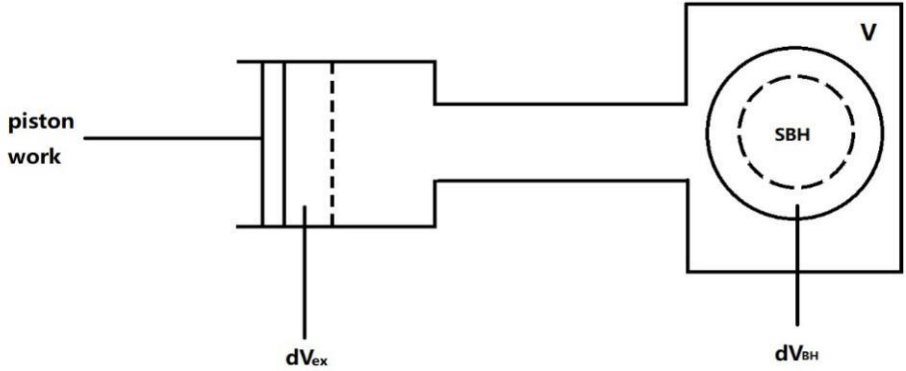

Figure 1. The illustration of a model for back reaction with a Schwarzschild black hole in a bulk volume surrounded by the radiation field connected to a piston to the outside on the edge.

## 2. Nonequilibrium surface tension

To consider the more general nonequilibrium cases, let us assume that the radiation field has the temperature $T_R$, and the black hole (horizon) has the temperature $T_{BH}$, while the temperature of the membrane is $T_H$. If we consider that the membrane is then and very close to the black hole surface, we can set $T_H \cong T_{BH}$. In this study, we consider near but away from the equilibrium case. We assume that the black hole, the membrane and radiation field have the same temperature at the beginning $T_{BH} = T_H = T_R$. As the time passes by, $T_{BH}$ and $T_R$ can become different (caused by the thermodynamic fluctuations and by black hole evaporation from Hawking radiation) but $T_H$ and $T_{BH}$ approximately keep the same, $T_H \cong T_{BH}$. We also assume that three subsystems (black hole and the membrane and the radiation field) are all at local equilibrium although the whole system is not at equilibrium. The irreversible processes caused by the heat or radiation transfer energy can emerge between the membrane and the radiation field. Therefore, one obtains:

$$dM_{BH} = T_{BH}dS_{BH} \quad (1)$$

$$dM_H = T_H dS_H + \sigma dA \quad (2)$$

$$dM_R = T_R dS_R - PdV \qquad (3)$$

Here, $M_{BH}$ represents the energy of the black hole, $M_H$ represents the energy of the back-reaction, and $M_R$ represents the energy of the radiation field. $P=\frac{1}{3}aT_R^4$, P represents the pressure of the black body radiation, $T_R$ represents the temperature of the radiation field, a represents a positive definite coefficient, and σ represents the equivalent tension coefficient on the membrane. (1) is derived from the first law of thermodynamics of black hole.[11]

In a large adiabatic bulk, the total changes of the energy (internal energy) of the whole system is given as:

$$d(M_{BH}+M_H+M_R) = -PdV_{ex} \qquad (4)$$

From (1)(2)(3), by using $T_H \cong T_{BH}$

$$PdV_{BH} = T_{BH}(dS_{BH} + dS_H + dS_R) - \Delta T dS_R + \sigma dA \qquad (5)$$

Here, $\Delta T = T_{BH} - T_R$. Assuming that the underlying process is adiabatic, then the total change of the reversible entropy is $(dS_{BH} + dS_H + dS_R) = 0$. Therefore, equation (5) turns into

$$\sigma = \frac{PdV_{BH}+\Delta T dS_R}{dA} = \frac{\frac{1}{3}aT_R^4 dV_{BH}+\frac{4}{3}aT_R^3 \Delta T dV}{dA}, \qquad (6)$$

Here, $dV=dV_{BH}+dV_{ex}$. Consider the situation that $dV_{ex}$ is far less than $dV_{BH}$. (It's reasonable as long as box is very large) In this way, we can omit $dV_{ex}$. We can see that in ordinary units

$$\sigma = \frac{-a\hbar c T_R^4}{24\pi k_B T_{BH}} - \frac{a\hbar c T_R^3 \Delta T}{6\pi k_B T_{BH}}. \qquad (7)$$

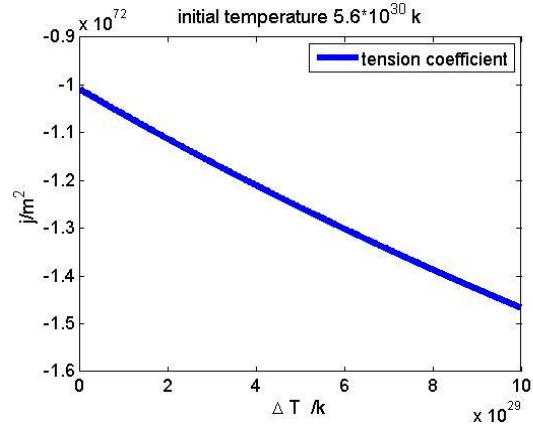

(2)

Figure (2) shows the relationship between the tension coefficient of the membrane of the black hole and temperature difference with an initial temperature $5.6\times 10^{30}$k. (for the rational we choose such a high temperature, see below)

Figure 2 shows the variation of the tension coefficient with respect to the temperature difference. Tension coefficient σ has a negative value. Tension coefficient of the membrane of the black hole decreases (the absolute value increases) as the temperature difference increases as shown in Figure 2. When $\Delta T = 0$, we recover the result of equilibrium state in [7].

We found that the larger the mass of the black hole, the smaller the absolute value of the surface tension coefficient, namely the weaker the back-reaction is. For various masses of black holes, the surface tension coefficients have the same trend as the temperature difference increases, σ always decreases as shown in Figure 2. The absolute value becomes larger, indicating that the larger the temperature difference is, the more significant the surface tension is and therefore more significant the effect of the membrane is, i.e. back-reaction is more prominent. The surface tension coefficient is negative. This indicates that if the area of the membrane becomes larger, the free energy of the membrane is reduced. Due to the existence of the temperature difference, the absolute value of the tension coefficient in the non-equilibrium scenario is significantly larger than that in the equilibrium state. While Hawking radiation has a tendency to shrink the area of the black hole horizon, the back-reaction has a tendency to enlarge the area. This might be why the surface tension coefficient is negative. The negative tension coefficient may also be expected due to the intrinsic gravitational interactions with negative heat capacity.

## 3. The reversible entropy changes of the membrane and the black hole against temperature difference

Differential expression of the free energy on the membrane is given as

$$dF_H = -S_H dT_{BH} + \sigma dA \quad (8)$$

$S_H$ is the entropy caused by the exchange of matter and energy between the membrane and the two sides. From (8), according to Maxwell relation, we can get

$$S_H = -\left(\frac{\partial \sigma}{\partial T_{BH}}\right)A = \frac{aT_R^4 \hbar^3 c^3}{32\pi^2 T_{BH}^4 k_B^3} \quad (9)$$

$S_H$ is the reversible entropy differentiating from $T_R$ (temperature of the black hole in equilibrium state) to $T_{BH}$ (temperature of the black hole at nonequilibrium state). Black holes of various masses have the same trend as the temperature difference becomes larger, and the reversible entropy gradually decreases. Because of the temperature difference, the back-reaction is distinctly different from the equilibrium case. The power emitted causes the mass to decrease at the rate $\frac{dM_{BH}}{dt} = -\frac{\hbar c^4}{G^2}\frac{\alpha}{M_{BH}^2}$, where $\alpha$ is a numerical coefficient. For $M_{BH} \gg 10^{17} g$, $\alpha = 2.001 \times 10^{-4}$, for $5 \times 10^{14} g \ll M_{BH} \ll 10^{17} g$, $\alpha = 3.6 \times 10^{-4}$ .[12] We can obtain $\dot{S}_{BH} = -\frac{8\pi\alpha k_B^2}{\hbar}T_{BH} \gg \dot{S}_H = -\frac{64\pi\alpha G a T_R^4 \hbar}{T_{BH}}$ .That means the reversible entropy change of the back-reaction is not significant compared to the black hole's, for a black hole with such mass. From (9), the total entropy of the black hole with the membrane (only consider reversible part) is $S_{dressed} = \frac{\hbar c^5}{16\pi G k_B T_{BH}^2}\left(1 + \frac{aT_R^4 \hbar^2 G}{2\pi c^2 T_{BH}^2 k_B^2}\right)$. If the back-reaction has a significant effect, it requires the mass of black hole to be around the magnitude of $\sqrt{\frac{\hbar^4 c^4 a}{128 G\pi^3 k_B^4}} \approx 1.4 \times 10^{-10} kg$. When the mass of black hole is much larger than $1.4 \times 10^{-10} kg$, the entropy of the black hole is much larger than the reversible entropy on the membrane, so the effect of back-reaction is weak. For a primordial black hole (PBH, suggested by Hawking [13]) with the minimum mass whose mass is $M_p = \sqrt{\frac{\hbar c}{G}} = 2.17 \times 10^{-8} kg$, which is formed in the early universe and evaporated rapidly over a very short life, its back-reaction is more prominent.

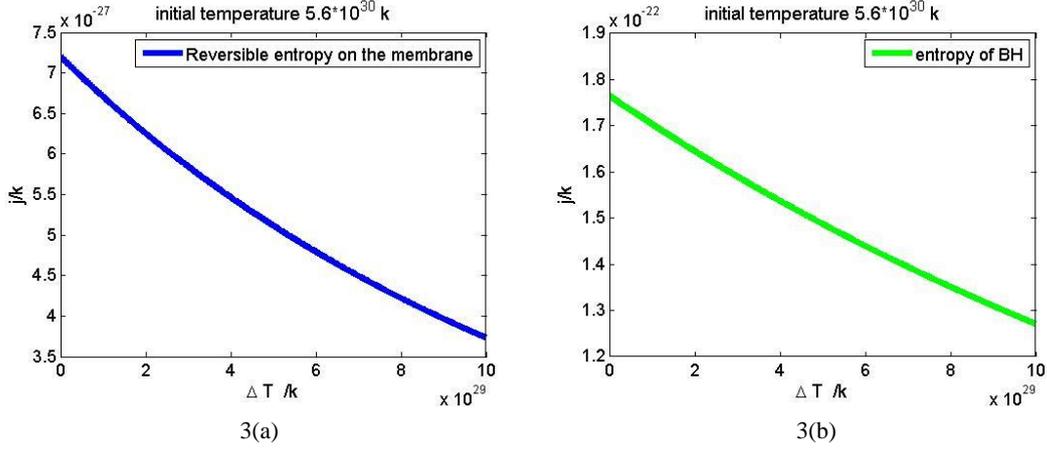

3(a)                                     3(b)

Figure 3: 3(a) shows the relationship between the reversible entropy of membrane and temperature difference with an initial temperature 5.6× $10^{30}$k (It was only in the very early universe that such high temperatures were possible). Black holes of various masses have the same trend as the temperature difference becomes larger, and the reversible entropy monotonically decreases. 3(b) shows change of entropy of the black hole with an initial temperature 5.6× $10^{30}$k.

Figure 3(a) shows that the reversible entropy of the membrane decreases as the temperature difference increases while Figure 3(b) shows that the entropy of the black hole decreases as the temperature difference increases. Thus the nonequilibrium effects characterized by the temperature difference can reduce the reversible entropy of the black hole. The back reaction of the membrane to the black hole on the entropy change is relatively small.

## 4. The energy changes of the membrane and the black hole against temperature difference

From equation (2), we can derive that

$$M_\text{H} = -\left(\frac{\partial \sigma}{\partial T_\text{BH}}\right) T_\text{BH} A + \sigma A = \frac{aT_R^4 \hbar^3 c^3}{48\pi^2 T_{BH}^3 k_B^3} - \frac{aT_R^3 \hbar^3 c^3 \Delta T}{24\pi^2 T_{BH}^3 k_B^3} \quad (11)$$

The energy of the black hole with the membrane is:

$$M_\text{H} + M_\text{BH} = \frac{aT_R^4 \hbar^3 c^3}{48\pi^2 T_{BH}^3 k_B^3} - \frac{aT_R^3 \hbar^3 c^3 \Delta T}{24\pi^2 T_{BH}^3 k_B^3} + \frac{c^5 \hbar}{8\pi G k_B T_{BH}} = \frac{c^5 \hbar}{8\pi G k_B T_{BH}}\left(1 + \frac{aT_R^4 \hbar^2 G}{6\pi c^2 T_{BH}^2 k_B^2} - \frac{aT_R^3 \hbar^2 G \Delta T}{3\pi c^2 T_{BH}^2 k_B^2}\right). \quad (12)$$

Figure 4(a) shows that the energy of the membrane monotonically decreases as the temperature difference increases. Figure 4(b) shows the energy of the black hole also decreases as the temperature difference increases. The nonequilibrium effect characterized by the temperature difference can reduce the reversible entropy and the energy of the black hole. The back reaction is relatively small compared to the black hole energy change.

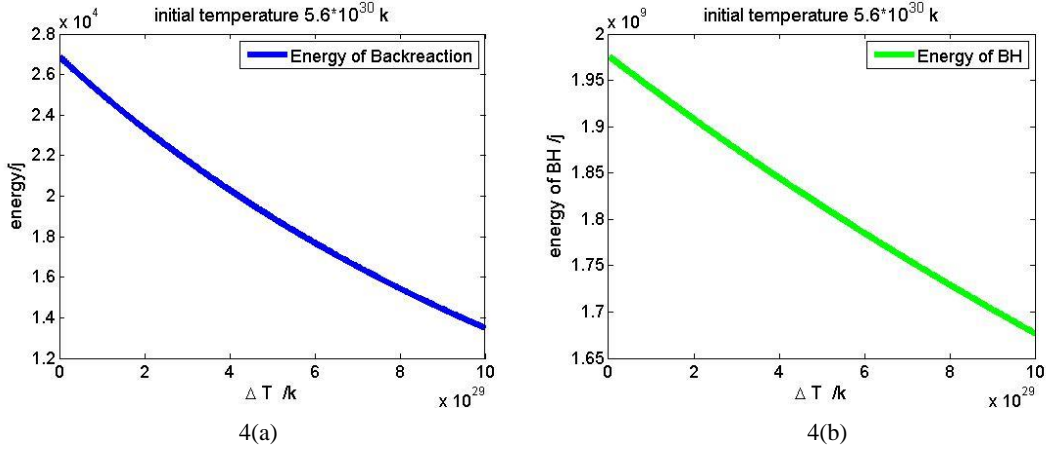

4(a)                                          4(b)

Figure 4(a) shows the energy of membrane monotonically decreases as the temperature difference increases. Figure 4(b) shows the relationship between the energy of the black hole and the temperature difference with an initial temperature $5.6\times 10^{30}$k.

It's easy to check if the back-reaction has a significant effect, it also requires mass of the black hole to be around the magnitude of $\sqrt{\frac{\hbar^4 c^4 a}{128 G \pi^3 k_B^4}} \approx 1.4 \times 10^{-10} kg$. We can conclude that only small mass primordial black hole can have a more significant back-reaction. The energy and reversible entropy of the back-reaction relative to the energy and entropy of the black holes can be measured by $\propto \frac{1}{M_{BH}^2}$, the more massive the black hole is, the weaker its back-reaction is. Moreover, in our nonequilibrium scenario, we can resolve the inconsistency issue of the black hole back reaction under zero mass limit appeared in the equilibrium scenario. For example, if we take a limit $M_{BH} \to 0$ (i.e. $T_{BH} \to \infty$ ), $M_{dressed}$ and $S_{dressed}$ will naturally vanish in our nonequilibrium model. We see that the nonequilibriumness can influence the evolution and back reaction of the black hole. Our nonequilibrium model is more natural and physically reasonable than the equilibrium model for the black hole back reaction and evolution.

## 5. The entropy production rate of the membrane and the rate of entropy change for the black hole against temperature difference

For a black hole with an ideal black body temperature $T_{BH}$ at the horizon immersed in a radiation field with a temperature $T_R$, the entropy production rate per unit surface area per unit time on the black body surface caused by temperature difference and the thermal radiation is given as

$\Theta = -\frac{4}{3}\omega T_R^3 + \frac{4}{3}\omega T_{BH}^3 + \frac{\omega T_R^4 - \omega T_{BH}^4}{T_{BH}}$, $\omega$ is the Stefan-Boltzmann constant.[14][15].

The entropy production rate formula above can be understood as the total entropy production from the system (the black hole) and environment (the radiation field). The last two terms represent the difference between the input radiation energy flux and the output black hole energy flux giving rise to the entropy production of the black hole. The first two terms represent the difference between the input black hole entropy flux and the output radiation field entropy flux giving rise to the entropy production of the radiation field. The Let us assume that the thermal radiation process relevant to the black hole only occurs near the horizon, and the entropy production of the surface of the black hole per unit time is $\frac{dS_i}{dt}=\Theta A$, where A is the area of the horizon. We can approximately estimate the increase of irreversible entropy. According to $T_{BH} = \frac{\hbar c^3}{8\pi G k_B M_{BH}}$ and A=16 $\pi \frac{G^2 M_{BH}^2}{c^4}$, Then we can obtain the nonequilibrium entropy production rate to be:

$$\frac{dS_i}{dt} = \frac{\hbar^2 c^2 \omega}{4\pi k_B^2 T_{BH}^2} \left(-\frac{4}{3}T_R^3 + \frac{4}{3}T_{BH}^3 + \frac{T_R^4 - T_{BH}^4}{T_{BH}}\right) \quad (10)$$

Black holes under various masses have the same trend of the changes as the temperature difference becomes larger. As the temperature difference becomes larger, the nonequilibrium thermodynamic dissipation quantified by the entropy production rate as well as the entropy change rate of the black hole increases significantly.

Importantly, we found that once there is a finite temperature difference, the additional entropy change caused by the radiation energy transfer of the total system in a unit time is significant as shown in Figure 5. Furthermore, as seen clearly in Figure 5, the nonequilibrium effect characterized by the temperature difference can enhance both the dissipation cost rate of the membrane and the entropy change rate of the black hole.

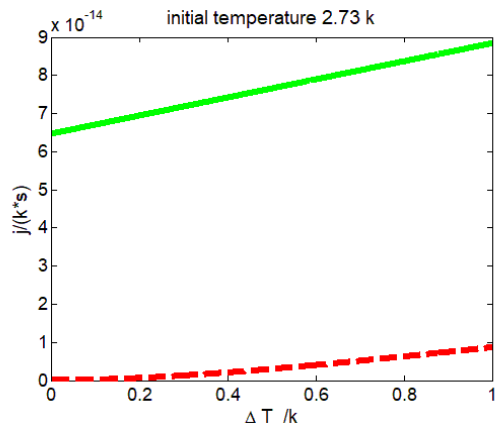

(5)

Figure 5 shows the variation of the irreversible entropy production rate (red line) and the rate of entropy change of black hole (green line) with temperature difference with an initial temperatures 2.73k. We find entropy production rate is significant compared to the rate of entropy change of a black hole at a finite temperature, it also holds for other black holes with different mass.

For the entropy change of the whole system, the effect of the irreversible nonequilibrium entropy production rate from radiation energy transfer appears to be significant compared to the rate of the entropy change of a black hole at a finite temperature difference as shown in Figure 5. The nonequilibrium entropy production rate can thus have an impact on the black hole evaporation rate. This may provide some insights towards the black hole information paradox since there can be significant information loss due to the entropy production rate in the nonequilibirum cases from the temperature difference between the black hole and the radiation environment.

# III. Nonequilibrium black hole thermodynamics and back reaction of the thick membrane

Now let us consider the case that the membrane is thick. When the membrane has a thickness, one has to take into account the changes in the temperature on the membrane, from the near black hole side at $T_{BH}$ where $T_{BH}$ is the black hole temperature to near radiation field side $T_R$ where $T_R$ is the environmental radiation field temperature surrounding the black hole and the back reaction membrane. If the membrane is not very thick, then it is reasonable to consider the situation where the temperature is linear in the length. The change in the temperature within the membrane is related only to the distance r to the black hole and not to the angle assuming isotropic for simplicity. For the system not too far from equilibrium and we can assume that the temperature within the membrane is linear in distance and can be expressed as $T=T_{BH}+(r-r_{BH})(T_R-T_{BH})/L$, where L is the thickness of the membrane, $r_{BH}$ is the black hole radius. We assume local equilibrium to be maintained as each layer of a sphere with radius r so that local equilibrium thermodynamics can apply, although the overall system is in nonequilibrium state.

**1. Nonequilibrium thermodynamics under local equilibrium assumption**

For a general thermodynamic system, the second law applies: $dS \geq \frac{dQ}{T}$, (dS is the entropy change of the system, dQ is the heat exchange between the system and the environment while T is the temperature.) which can be generalized to the open system as $dS = d_e S + d_i S$, where $d_e S$ represents the change in entropy of the exchange of matter and energy between the system and the environment while $d_i S$ represents the change in entropy caused by the irreversible processes within the system. Let us consider the following situation. The overall system is in nonequilibrium state while its small subsystems is still in local equilibrium and the equilibrium thermodynamic laws still hold for the subsystems, $Tds = du$, where $s$, $T$, $u$, are entropy, temperature, and energy of the subsystem respectively. Therefore, for the whole system, $S = \int s dV, U = \int u dV$.

In the simple process of heat transport, the local change of the internal energy is equal to the thermal heat flux in and out of the subsystem: $\frac{\partial u}{\partial t} = -\nabla \cdot \boldsymbol{J}_q$, where $\boldsymbol{J}_q$ is the heat flow per unit area per unit time. Therefore, $\frac{\partial s}{\partial t} = \frac{1}{T}\frac{\partial u}{\partial t} = -\frac{1}{T}\nabla \cdot \boldsymbol{J}_q = -\nabla \cdot \frac{\boldsymbol{J}_q}{T} + \boldsymbol{J}_q \cdot \nabla \frac{1}{T}$. In the general cases, under local equilibrium for the subsystems, the entropy change rate is determined by $\frac{\partial s}{\partial t} = -\nabla \cdot \boldsymbol{J}_s + \odot$ where $\boldsymbol{J}_s$ is the entropy flow per unit area per unit time, and $\odot$ is entropy production rate per unit volume [16]. The entropy change rate of the whole system is given as $\frac{dS}{dt} = \frac{d}{dt}\int s dV = \int \frac{\partial s}{\partial t} dV = \int(-\nabla \cdot \boldsymbol{J}_s + \odot)dV = -\oint \boldsymbol{J}_s dA + \int \odot dV$. The first term is the entropy flow in from the outside through the surface of the system per unit time. The second term is the sum of the entropy production of all the subsystems per unit time. Comparing the heat transport case with the general entropy change formula under local equilibrium assumption with $\frac{\partial s}{\partial t} = -\nabla \cdot \boldsymbol{J}_s + \odot$, we can see that $\boldsymbol{J}_s = \frac{\boldsymbol{J}_q}{T}$ and $\odot = \boldsymbol{J}_q \cdot \nabla \frac{1}{T}$.

## 2. Nonequilibrium black hole thermodynamics under local equilibrium assumption

In our model, for both ends of the membrane, the input energy flows per unit time and per unit area are given as $J_{BH} = \omega T_{BH}^4$ from the black hole and $J_R = \omega T_R^4$ from the radiation field, $\omega$ is the Stefan-Boltzmann constant. The outward radiation along the radius is in the positive direction. The radiation energy flow in the membrane is $J_q = \omega T_{BH}^4 - \omega T_R^4$. Although we assume the local equilibrium at each layer of the radius r, the system overall is in nonequilibrium state with temperature distribution varying in space. The energy flows are assumed mostly through the radiation rather than thermal conduction in our study.

The rate of the exchange in entropy between the back reaction membrane system and the environment per unit area in the membrane is given as $\dot{s}_e = -\nabla \cdot \frac{J_q}{T} = \frac{J_q}{T^2}\nabla T = J_q \frac{\Delta T}{T^2 L}$, where $\nabla T = \frac{dT}{dr} = \frac{\Delta T}{L}$ is from our setting T=$T_{BH}$+(r-$r_{BH}$)($T_R$−$T_{BH}$)/L and $\Delta T = T_R - T_{BH}$ here. The radiation energy flow in the membrane $J_q = \omega T_{BH}^4 -$

$\omega T_R^4$ is constant analogous to a case of an electric current $J_q$ moving through a resistance of length L under a voltage is $\nabla \frac{1}{T}$. The entropy production rate per unit area is given as $\dot{s}_i = \Theta = J_q \cdot \nabla \frac{1}{T} = -J_q \frac{\Delta T}{T^2 L} > 0$. Since the energy flux $J_q$ is a constant in the membrane, we can check that the system entropy of the membrane does not change and is therefore in steady state. Since the total entropy production rate per unit area of the membrane system and the environment is the sum of the system entropy is equal to the entropy change rate of the system and its exchange with the environments (same as the environmental entropy change) per unit area according to the nonequilibirum thermodynamics, the entropy production rate is equal to the entropy exchange rate with the environment or the environmental entropy change rate. For the whole membrane system, we can obtain the total entropy exchange between the membrane system and the environments as $\dot{S}_e = \int \dot{s}_e dV = \int \dot{s}_e \, 4\pi r^2 dr$. Suppose the thickness of the membrane is much less than the radius of black hole, $L \ll r_{BH}$. Then we can obtain approximately $\dot{S}_e \approx A_{BH} \int \dot{s}_e \, dr = J_q A_{BH} \left( \frac{1}{T_{BH}} - \frac{1}{T_R} \right) = \frac{J_q \hbar^2 c^2}{4\pi T_{BH}^2 k_B^2} \left( \frac{1}{T_{BH}} - \frac{1}{T_R} \right)$ where the area of the black hole is given as $A_{BH} = \frac{\hbar^2 c^2}{4\pi T_{BH}^2 k_B^2}$. For a black hole at temperature 2.7k, its radius is given as $r_{BH} = \frac{\hbar c}{4\pi T_{BH} k_B} = 6.68 \times 10^{-5} m$, so the thickness of the membrane should be less than $10^{-7} m$ at least such that $4\pi (r_{BH} + L)^2 = 4\pi r_{BH}^2 + 4\pi L^2 + 8\pi r_{BH} L \approx 4\pi r_{BH}^2$. Also, the total entropy production rate of the membrane system is approximately given as $\dot{S}_i = \int \dot{s}_i \, dV \approx \frac{J_q \hbar^2 c^2}{4\pi T_{BH}^2 k_B^2} \left( \frac{1}{T_R} - \frac{1}{T_{BH}} \right) > 0$.

## 3. The rate of energy and entropy change of the back reaction on the black hole against temperature difference

We now explore how the entropy change of the black hole, the energy change, the mass change of the black hole, and the mass or energy of the membrane change with respect to time, $\dot{S}_{BH} = -\omega T_{BH}^3 A_{BH} = -\frac{\omega \hbar^2 c^2 T_{BH}}{4\pi k_B^2}$ where $\omega T_{BH}^3$ is the entropy flow from black hole and $\dot{M}_{BH} = -J_{BH} A_{BH} = -\frac{\omega \hbar^2 c^2 T_{BH}^2}{4\pi k_B^2}$, $\dot{M}_H = -(\omega T_{BH}^4 A_{BH} - \omega T_R^4 \times 4\pi (r_{BH} + L)^2)$, where the energy flux $J_{BH} = \omega T_{BH}^4$ is from the black hole and the energy flux $J_R = \omega T_R^4$ is from the radiation field. Using the assumption that the thickness of the membrane is much less than the black hole radius, we obtain $\dot{M}_H \approx -(\frac{\omega T_{BH}^2}{4\pi} - \frac{\omega T_R^4}{4\pi T_{BH}^2}) \frac{\hbar^2 c^2}{k_B^2}$.

Let us assume that the expression of $\dot{S}_{BH}$ and $\dot{M}_{BH}$ above are approximately valid for the 'naked' black hole, and the back reaction of the entropy rate and energy rate from the membrane are given as $\dot{S}_e$ and $\dot{M}_H$ respectively. After considering the back reaction, the effective or dressed entropy rate and energy rate of the black hole are given as $\dot{S}_{dressed} = \dot{S}_H + \dot{S}_{BH}$, $\dot{M}_{dressed} = \dot{M}_H + \dot{M}_{BH}$ respectively.

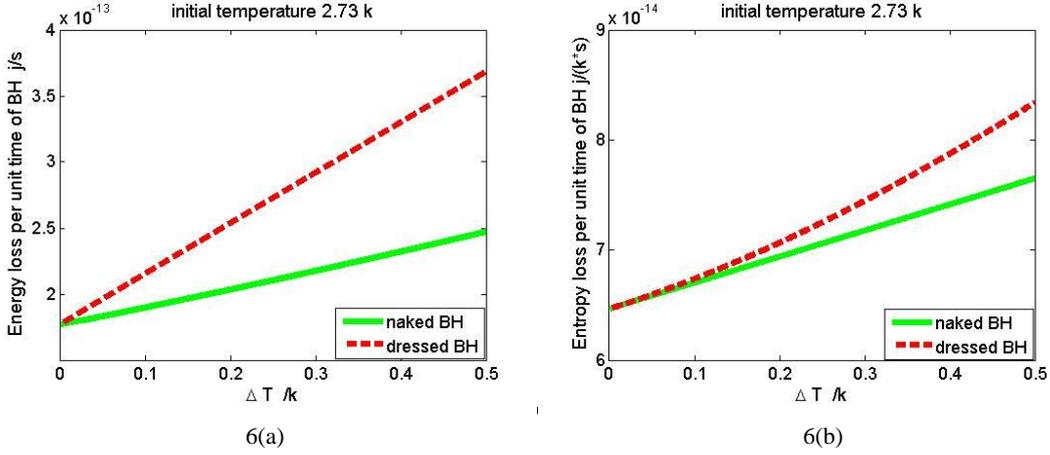

6(a)            6(b)

Figure 6(a) 6(b) : The energy and entropy loss per unit time of 'naked' black hole and 'dressed' black hole varying with the temperature difference with an initial temperature 2.73 k. Here $\Delta T = T_{BH} - T_R$.

When $\Delta T = 0$, the energy and entropy loss per unit time of 'dressed' black hole is equal to that of the 'naked' black hole's as shown in Figure 6(a) and 6(b). The system is in equilibrium. When the temperature difference is larger, the system becomes more nonequilibrium. Compared to the thin membrane case, we find that the thick membrane can lead to more significant back reactions on the black holes. In both Figure 6(a) and 6(b), we see that the energy and entropy change rate of a dressed black hole after taking into account of the back reaction are larger than that of the 'naked' black hole', and the distances between the green line and red line increase as $\Delta T$ increases. This shows that the back-reaction on the energy rate and entropy rate of the black hole is more significant as the temperature difference increases. The evaporation rate of the dressed black hole when taking into account of the thickness of the membrane back reaction is larger than that of the 'naked black hole'. On the other hand, from both Figure 6(a) and 6(b), the nonequilibrium effects characterized by the temperature difference can enhance the energy and entropy loss rate of the black hole and the membrane back reaction.

## 4. Energy and entropy change of the black hole due to the back reaction against temperature difference

We would not only like to see how significant the rate of the back reaction but also the back reaction of the membrane itself is on the black hole. To do so, we need to compare the contribution of the entropy and energy from the membrane towards the black hole. Let us estimate the time scale given the entropy and energy rate of the membrane above. The response of a black hole to a perturbation will be dominated by a set of damped oscillations called quasi-normal modes. The quasi-normal mode frequency corresponds to a natural frequency of a black hole. It is divided into two parts: the real part represents the oscillation and the imaginary part represents damping [17]. For large damping, i.e., for a large imaginary part of ω, the real part of the frequency approaches a nonzero value and the imaginary part becomes equally spaced. It is given as $M\omega = 0.04371235 + i(n + \frac{1}{2})$ [18]. The constant real part of the quasi-normal frequency can be approximated as $w_{QNM} = \frac{\ln 3\, c^3}{8\pi GM}$. Thus, the natural time scale can be estimated as $\tau = \frac{2\pi}{w_{QNM}} = \frac{16\pi^2 GM}{\ln 3\, c^3} = \frac{2\pi \hbar}{\ln 3\, k_B T_{BH}}$.

In the time interval $\tau$, $\Delta S_e = \dot{S}_e \tau = \frac{J_q \hbar^3 c^2}{2\ln 3\, k_B^3 T_{BH}^3}(\frac{1}{T_{BH}} - \frac{1}{T_R})$, $\Delta S_i = \frac{J_q \hbar^3 c^2}{2\ln 3\, k_B^3 T_{BH}^3}(\frac{1}{T_R} - \frac{1}{T_{BH}})$, $\Delta S_{BH} = -\frac{\omega \hbar^3 c^2}{2\ln 3\, k_B^3}$, $\Delta M_{BH} = -\frac{\omega \hbar^3 c^2 T_{BH}}{2\ln 3\, k_B^3}$, $\Delta M_H = -(\frac{\omega T_{BH}}{2} - \frac{\omega T_R^4}{2T_{BH}^3})\frac{\hbar^3 c^2}{\ln 3\, k_B^3}$. We measure the back-reaction as $\frac{\Delta S_e}{\Delta S_{BH}} = (T_{BH}^4 - T_R^4)(\frac{1}{T_{BH}T_R^3} - \frac{1}{T_{BH}^4})$ and $\frac{\Delta M_H}{\Delta M_{BH}} = 1 - T_R^4/T_{BH}^4$. In the case of thick membrane, the energy and entropy change of back-reaction during certain periods is dependent on the temperature difference.

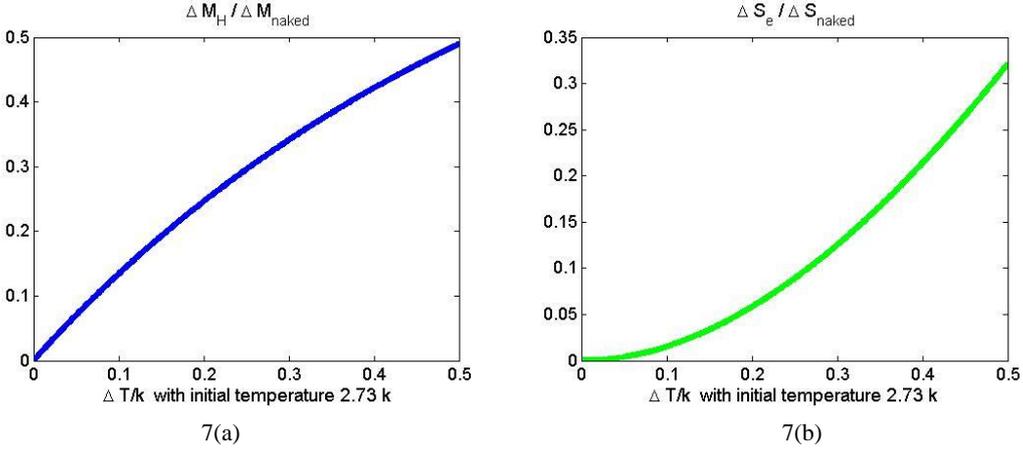

7(a)   7(b)

Figures 7(a) 7(b) show $\frac{\Delta M_H}{\Delta M_{BH}}$, $\frac{\Delta S_e}{\Delta S_{BH}}$ varying with the temperature difference with an initial temperature 2.73 k respectively.

When $\Delta T = 0$, the system is in equilibrium, $\Delta M_H = \Delta S_e = 0$ in both 7(a) and 7(b). According to figures 7(a) and 7(b), both $\frac{\Delta M_H}{\Delta M_{BH}}$ and $\frac{\Delta S_e}{\Delta S_{BH}}$ increase as the temperature difference increases. The back reaction becomes more significant as temperature difference increases. Also, for a black hole with a mass $M_{BH} \gg 5 \times 10^{14}$g, the effect of thick membrane ($\dot{S}_e = \frac{J_q \hbar^2 c^2}{4\pi T_{BH}^2 k_B^2}(\frac{1}{T_{BH}} - \frac{1}{T_R})$) is much more significant than that of thin membrane ($\dot{S}_H = -\frac{\omega^2 \hbar^4 G T_R^4}{4\pi^2 k_B^4 T_{BH} c}$) under finite temperature difference. Under an infinitesimally small temperature difference, the effect of a thick membrane is similar to that of a thin membrane, they are both weak. Therefore, we can conclude that the back reaction from the membrane to the black hole can be significant in terms of the energy (mass) and entropy change. The nonequilibrium effect characterized by the temperature difference can enhance the back reactions in terms of the energy (mass) and entropy change to the black hole.

## 5. Nonequilibrium entropy production and entropy change of the black hole against temperature difference

For irreversible nonequilibrium entropy production, once there is a finite temperature difference, the additional entropy change caused by the radiation energy transfer of the total system in a unit time can be significant as shown in 8(a). It can be quantified by $\frac{\Delta S_i}{\Delta S_{BH}} = (T_{BH}^4 - T_R^4)(\frac{1}{T_{BH}T_R^3} - \frac{1}{T_{BH}^4})$.

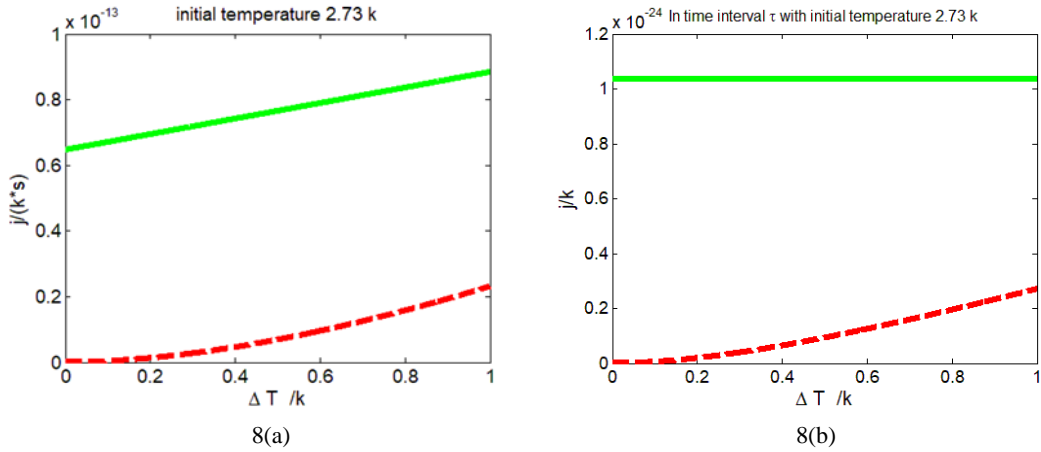

| 8(a) | 8(b) |

Figure 8(a) shows the variation of the irreversible entropy production rate (red line) and the rate of entropy change of the black hole (green line) with respect to the temperature difference with an initial temperature 2.73k. Figure 8(b) shows the entropy change caused by irreversible entropy production (red line) and the entropy change of the black hole caused by radiation (green line) against the temperature difference with an initial temperature 2.73k in time interval τ.

Figure 8(a) shows the variation of the irreversible entropy production rate (red line) and the rate of entropy change of black hole (green line) with respect to the temperature difference at initial temperatures 2.73k. We found as the temperature difference increases, both the entropy production rate and rate of entropy change of the black hole increase. This indicates that the nonequilibrium effects characterized by the temperature difference can increase the dissipation cost rate and entropy change rate of the black hole. We also find that the entropy production rate is significant compared to the rate of entropy change of black hole at a finite temperature difference. This also holds for black holes with different masses. Figure 8(b) shows the entropy change caused by irreversible entropy production rate (red line) and the entropy change of black hole caused by radiation (green line) with the temperature difference at initial temperatures 2.73k in time interval τ. Notice that the green line in Figure 8(b) is a constant. This is due to the underlying assumption that the entropy (area) of a black hole radiating out in time interval τ is quantized from the quasi-normal mode frequency expression [19]. This shows τ as a natural time of black hole. We found as the temperature difference increases, both the entropy production and entropy change of the black hole increase. This shows that the nonequilibrium effects characterized by the temperature difference can increase the dissipation cost and entropy change of the black hole. The entropy change caused by irreversible entropy production is significant compared to the entropy change of a black hole caused by radiation at a finite temperature difference. This also holds for black holes with different mass. The nonequilibrium entropy production rate can thus have an impact on the black hole evaporation rate.

This indicates that there can be significant entropy or information loss under the nonequilibrium conditions characterized by the temperature difference between the black hole and the radiation environment. This extra loss of the entropy or information may give some insights towards the understanding the source of the black hole information paradox.

## IV. Nonequilibrium thermodynamic fluctuations of the black hole and back reaction

The power emitted causes the mass of black hole to decrease at the rate $\frac{dM_{BH}}{dt} = -\frac{\hbar c^4}{G^2}\frac{\alpha}{M_{BH}^2}$. So, if the black hole is massive enough, i.e., its mass exceeds $5 \times 10^{14} g$, the loss rate of these quantities is almost constant and the evaporation process may be approximated by a nonequilibrium quasi steady-state, provided that the time scale of any realistic measurement is much shorter than the time scale of the process($\propto M_{BH}^3$). [20] It is easy to check that for $M_{BH} > 10^{14}g$, the characteristic decay time is of the order of the age of the Universe. The electromagnetic radiation and its back reaction on the metric were considered in a stochastic Langevin approach [21]. The results yielded a lower limit on the statistical variations of the metric itself, which could be thought as the 'seismology' of the spacetime's geometry, as a consequence of the back reaction. In this spirit, one can consider the stochastic equation $\dot{M}_{BH} = \frac{-\alpha}{M_{BH}^2} + \delta\dot{M}$, where $\delta\dot{M}$ represents a spontaneous thermal fluctuation of $\dot{M}_{BH}$, such a thermal fluctuation in the black-hole should be reflected in the fluctuations of the background metrics and thereby may be regarded as reflecting the effects of back reaction of the Hawking radiation to the evolution of a black hole evaporation within a phenomenological level.[21][22]

By using Landau-Lifshitz theory[23], for a general fluctuating dissipative system, $\dot{x}_i = -\sum_j \Gamma_{ij} X_j + \delta\dot{x}$, $\dot{x}_i$ is a general flux and $X_i$ is the thermodynamic force conjugate to the flux $\dot{x}_i$, $\Gamma_{ij}$ is phenomenological transport coefficients which is defined as $\Gamma_{ij} = \frac{\partial \dot{x}_i}{\partial X_j}$ and $\delta \dot{x}$ stands for fluctuations. Entropy production rate can be given as $\dot{S} = -\sum_i \dot{x}_i X_i$ from nonequilibrium thermodynamics. Because the entropy of the Schwarzschild black hole is not concave, the above expression must be modified as $\dot{S} = \sum_i \pm(\dot{x}_i X_i)$. (for more details see Ref [20]). Therefore, for Schwarzschild black hole, $\dot{S} = \frac{\dot{M}_{BH}}{T_{BH}} = \frac{-8\pi\alpha}{M_{BH}}$, $\dot{M}_{BH}$ can be interpreted as the flux or current, $\frac{1}{T_{BH}}$ can be interpreted as the thermodynamic driving force conjugate to the flux $\dot{M}_{BH}$. Then the second order moments in the fluctuations of the fluxes can be expressed as $\langle \delta \dot{x}_i \delta \dot{x}_j \rangle = (\Gamma_{ij} + \Gamma_{ji})\delta_{ij}$.

Now one can derive that $\langle \delta \dot{M} \, \delta \dot{M} \rangle = \alpha(4\pi M_{BH}^3)^{-1}$ from $\Gamma_{MM} = \frac{\partial \dot{M}_{BH}}{\partial \frac{1}{T_{BH}}}$ and $\langle \delta \dot{S} \delta \dot{S} \rangle = (\frac{\partial S}{\partial M_{BH}})^2 \langle \delta \dot{M} \, \delta \dot{M} \rangle = 16\pi\alpha M_{BH}^{-1}$, where $S = 4\pi M_{BH}^2$. The second order moments reflect the fluctuation intensity of the black hole, namely, the intensity of the back-reaction. Therefore, we can also conclude that the more massive the black hole is, the weaker its back-reaction is. This argument holds for a black hole with mass exceeding $5 \times 10^{14} g$, since we can consider the evaporation of the black hole as a nonequilibrium quasi steady-state process only in this situation.

## V. Conclusion

In this study, we developed a nonequilibrium thermodynamics model for investigating the back reaction of the Hawking radiation on the Schwarzschild black hole. In our study, two scenarios are considered: the thin membrane and the thick membrane. The thin membrane formed has a negative tension coefficient. We found that larger temperature difference can lead to more significant negative surface tension and back reaction as well as larger black hole area. The thin membrane has the tendency to expand to counteract the tendency of Hawking radiation for reducing the black hole area. We showed that the more massive the black hole is, the weaker its back reaction is based on a nonequilibrium thermodynamic approach of the thin membrane. We also confirm this using a phenomenological fluctuation analysis. We conclude that at the non-equilibrium state, the back-reaction may contribute to the change of the entropy and energy for the primordial black hole. The larger temperature difference between the black hole and the radiation environment gives rise to more significant back-reaction. Moreover, our nonequilibrium model can resolve the inconsistency issue of the black hole back reaction under zero mass limit in the equilibrium case. Our nonequilibrium model appears to be more natural and physically reasonable than the equilibrium model for the black hole back reaction and evolution. We also study the thick membrane assuming the thickness of the membrane is much less than the radius of black hole such that the temperature can be viewed to vary uniformly in the membrane. We find that the thick membrane shows more significant effects on back reactions compared to the thin membrane.

Importantly, we also quantified the nonequilibrium thermodynamic dissipation represented by the entropy production rate. We showed that the entropy production rate increases significantly as the temperature difference increases for both the thin membrane and the thick membrane. We found that in general the nonequilibrium effects characterized by the temperature difference can increase the energy and entropy loss as well as the entropy production rate of the black hole and the membrane back reactions. We show that once there is a finite temperature difference the nonequilibrium entropy production rate appears to be significant compared to the entropy rate radiated by the black hole. The nonequilibrium entropy production rate can thus have an impact on the black hole evaporation rate. This may have impacts on the black hole information paradox due to the significant information loss from the entropy production rate under nonequilibrium scenarios (temperature difference between the black hole and the surrounding radiation environment).

## Data Availability

The data used to support the findings of this study are included within the article.

## Conflicts of Interest

The authors declare that they have no conflicts of interest.